\documentclass[slac_one]{revtex4}
\usepackage{epsfig}
\usepackage{graphicx}
\usepackage{fancyhdr}
\usepackage{amsmath}
\usepackage{amssymb}
\newcommand\bea{\begin{eqnarray}}
\newcommand\eea{\end{eqnarray}}

\newcommand\as{\alpha_s}

\begin{document}

%Title of paper
\title{
Review of Theoretical Status:\\
the Long and Short of High Energy Jets\\}

% Repeat the \author .. \affiliation  etc. as needed
%
% \affiliation command applies to all authors since the last
% \affiliation command. The \affiliation command should follow the
% other information

\author{George Sterman}
%%%MY ADDRESS
\affiliation{C.N.\ Yang Institute for Theoretical Physics, Stony Brook University,
Stony Brook,, NY 11794-3840 USA }

\begin{abstract}
High energy jets and their associated event shapes provide a window
into the transition between elementary and composite degrees of freedom
in quantum chromodynamics.   This talk reviews some methods 
and a few principles that have led to progress in this area.\end{abstract}

%\maketitle must follow title, authors, abstract
\maketitle

\thispagestyle{fancy}

% body of paper here - Use proper section commands
% References should be done using the \cite, \ref, and \label commands
% Put \label in argument of \section for cross-referencing
%\section{\label{}}

\section{Motivation: QCD, the Protean Theory}

Why fuss over quantum chromodynamics?  Why not simply
leave a true theory alone, except perhaps to derive it
from some underlying physics beyond the standard model?
The Large Hadron Collider, of course, plans to put QCD
to work in coaxing new states out of virtuality and into plain sight. 
Another enduring challenge in contemporary science, however,
is  to bridge the gap between a reductionist
description of nature in ``elementary" terms,  and an effective
description in terms of ``emergent" excitations.
Quantum chromodynamics
is a perfect testing ground for such a program.   In the
words of Shakespeare's Duke of  Gloucester, it
``can add colors to the chamelion/Change shapes with Proteus 
for advantages" [Henry VI, part III], and out it emerge
a host of effective theories at varying length scales.  
Quantum chromodynamics spans the chasm between the theory of
Yang and Mills at high energies and that of  Yukawa
at low energies, not to mention nuclei, and the description of 
strongly interacting matter at
varying temperatures and densities.
The only problem is that these aspects of the theory
can very difficult to analyze with
our current tools.

%\subsection{The Beginning of this era and today}

The current era of high energy colliders brought
the theory of QCD into being.   This required a temporary step
back  from the study of the fine structure of final states,
to their gross properties, which can encode events at short distances.
Before this was realized, however, the detailed
structure of final states were front and center.  In this context,
 Yang observed in 1969 that \cite{yang69}
 ``\dots  in the midst of enormous complexities
there are  very striking 
characteristics exhibited by high energy collisions
\dots A number of ideas, models etc. 
 been introduced \dots ideas behind these  models are
not always mutually exclusive, especially since none of
them is completely precisely defined."  
In many ways, this describes our current situation.
Now we have a theory, QCD, and we are ready to
turn our attention back to the individual particles 
produced by the strong interactions.
We understand some of these regularities as keys
to the protean transition from weak to strong coupling.
 Our challenge is to use these characteristics as guides 
for progress from models to principles.  
We have a very long way to go before we can
truly bridge this gap, but we have made some
halting progress.  And in a sense that is the subject
of our workshop.

\section{Power Corrections from Perturbation Theory: the OPE and a Massive Gluon}

We begin with a compressed intellectual history of 
observations that make it possible to derive power
corrections from perturbation theory.  This subject is intimately related to
the evolution of the QCD coupling.  It begins most naturally with 
fully inclusive cross sections, which can be
related to Green functions in Euclidean space.

\subsection{Discovering the operator expansions}

\begin{figure}[t]
\epsfig{figure=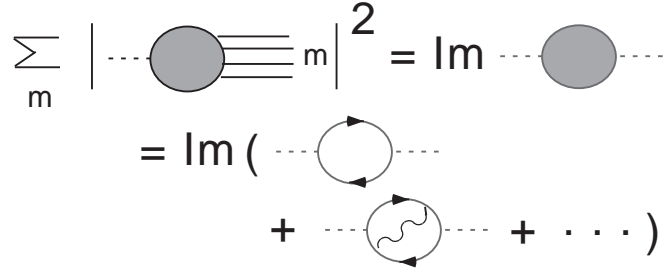,width=0.5\textwidth}
\caption{Graphical representation of the optical theorem as applied to
leptonic annihilation through a vector boson.\label{optical}}
\end{figure}

Our discussion begins with the work of
Mueller \cite{mueller85}, in 1985.  't Hooft \cite{thooft74} had previously introduced the concept
of renormalons, applied to Euclidean Green functions, but it was
Al who made the crucial step into Minkowski space, by 
exploiting the unitarity relation 
\bea
\sigma_{\rm tot}^{{\rm e}^+{\rm e}^-}
={(4\pi \alpha)^2\over Q^2}\, {\rm Im}\; \pi(Q^2)
\eea
between the total cross section
for lepton-positron annihilation 
and the imaginary part of
the time-ordered products of currents
\bea
\pi(Q^2)=\left({-i\over 3Q^2}\right)\; \int d^4x\; {\rm e}^{-iq\cdot x}
\langle 0|T j^\mu(0)j_\mu(x)|0\rangle\, .
\eea
This is the optical theorem, illustrated in Fig.\ \ref{optical}.  
It provides a link between a short-distance dominated
quantity, the self-energy of an  off-shell vector boson
(photon, Z) and the annihilation cross section, which of course is
the (inclusive) result of measurements at arbitrarily large times.

The important relationship  between the operator product expansion
for the product of currents
\bea
\langle 0|j^\mu(0)j_\mu(x)|0\rangle 
= {1\over x^6}\; {\tilde C}_0(x^2\mu^2),\as(\mu)) 
\ +{1\over x^2}\; {\tilde C}_{F^2}(x^2\mu^2,\alpha_s(\mu))
\langle 0|F_{\mu\nu}F^{\mu\nu}(0)|0\rangle +\dots\, ,
\label{ope}
\eea
and the total cross section was already
well-established.  In the method of QCD sum rules \cite{qcdsum}, the
first term of this expansion is interpreted as finite-order perturbation
theory, while additional terms are computed by introducing
new parameters, the condensates, in an extension of the 
perturbative series.
The consistency between $\pi(Q)$ computed in this way
and experiment leads to many predictions, always keeping
in mind that the values of the nonperturbative parameters
depend on the order of perturbation theory.

On the other hand, if we stick to perturbation theory at the outset, we find that
the perturbative expansion for the function $\pi(Q)$ is completely well-defined
and finite order-by-order, a property called infrared safety:
\bea
\pi(Q) \sim C_0(Q^2/\mu^2,\alpha_s(\mu))=
\sum_{n=0}^\infty c_0^{(n)}(Q^2/\mu^2)\; \alpha_s^n(\mu)\, ,
\eea
with $\mu$ the renormalization mass.
It would seem then, that perturbation theory knows only about the
leading power coefficient, $C_0$.
So what about $\tilde C_{F^2}$ and the rest of the coefficients?
In fact, although the other terms are not predicted by QCD perturbation theory,
their presence is implicit in it.
The reason is that the series for $C_0$ doesn't converge.

\begin{figure}
\epsfig{figure=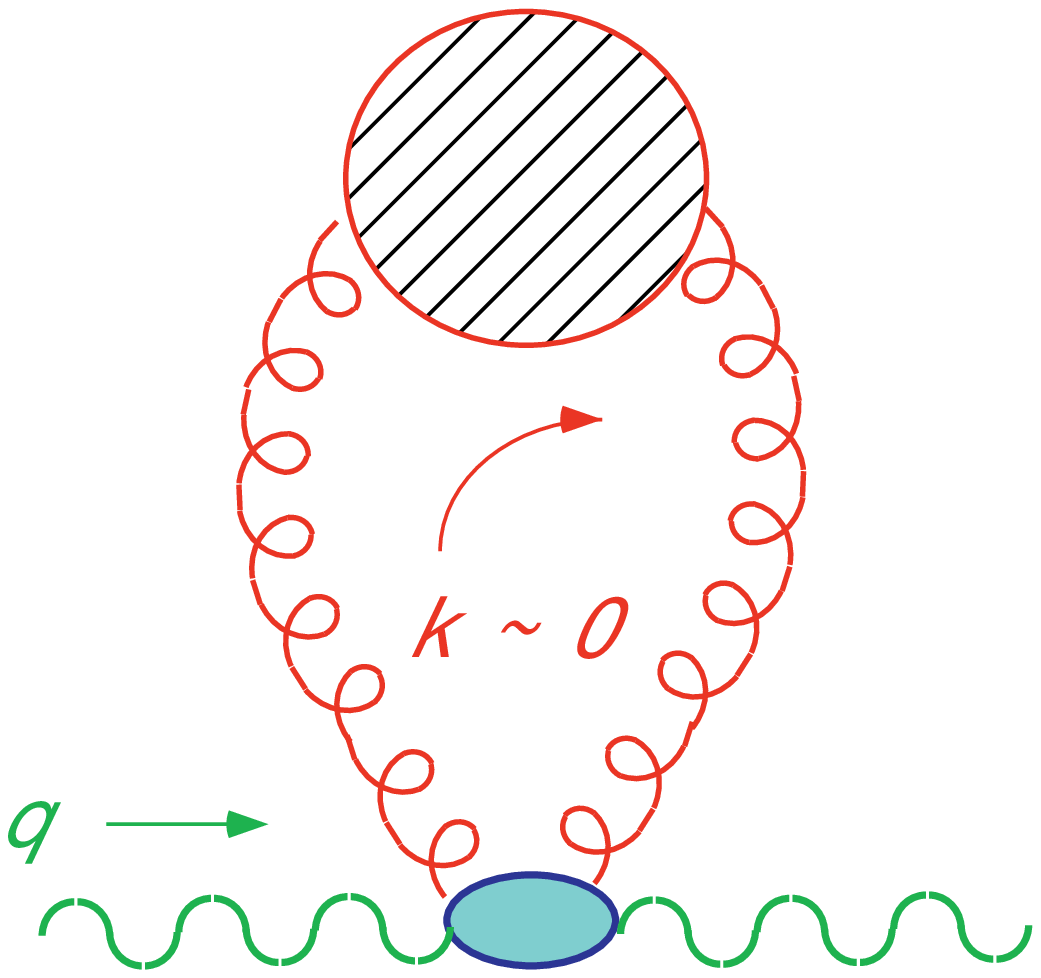,width=0.3\textwidth}
\hfil
\epsfig{figure=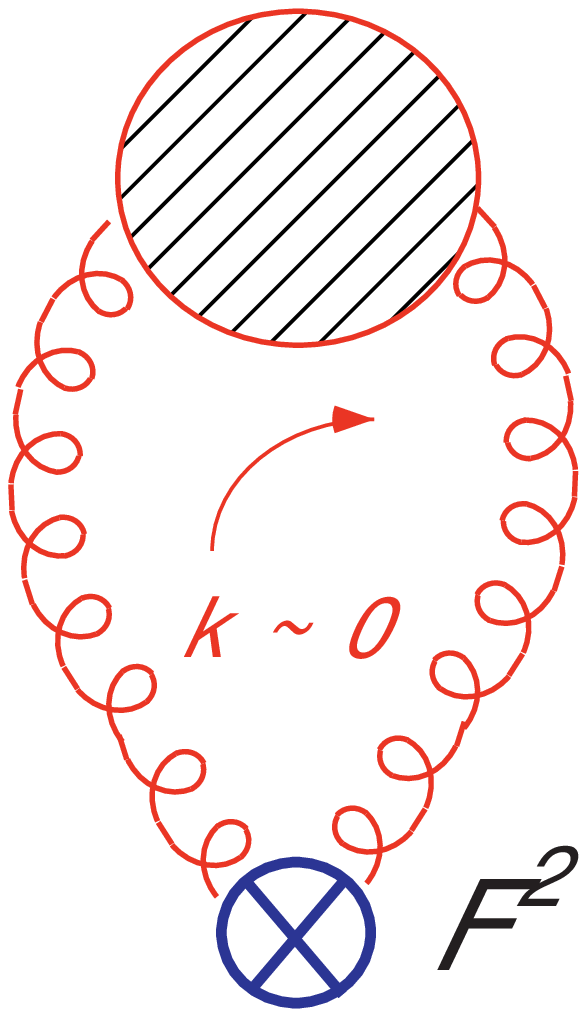,width=0.18\textwidth}
\hfil
\vskip 8mm

\centerline{(a) \hbox{\hskip 7 cm (b)}}
\caption{Momentum configuration related to renormalons
and the operator product expansion. \label{opepinch}}
\end{figure}

To see how this nonconvergence comes about, we can consider
a point in the loop momentum space of any diagram for $\pi(Q)$,
where the momentum $k$ carried by some gluon vanishes
in all four components.  In the language of \cite{gs78},
the subspace where this happens is a ``pinch surface",
which cannot be avoided by deforming the contours of loop momenta.  
The configuration is illustrated in Fig.\ \ref{opepinch}.

As we integrate over the loop momentum $k$, the invariant 
$k^2$  varies.  As we shall see, it is sometimes convenient
to think of this as a ``gluon mass".  In the figure we show
a gluon self-energy, which in general includes a series
of one-particle irreducible diagrams.  Each such diagram
behaves at lowest order as a (negative) constant times 
$\as\, \ln(k^2/\mu^2)$.
Thus, a typical term  in the perturbative expansion of the gluon
self energy produces a
$k$ integral that may be approximated for $0<k < Q^2$ by
\bea
C_0 &\sim&(1/Q^4)\; \as^n\, \int_0^{Q^2} dk^2 k^2\; \ln^n(k^2/\mu^2) 
\sim
(1/2)\left(\as/2\right)^n\; n!\, .
\eea
Dimensional counting plus the requirements of gauge invariance
ensure that the integrand vanishes as $k^2/Q^2$ for $k^2\to 0$, which
corresponds dimensionally to the operator $F^2$ in Eq.\ (\ref{ope}).  Indeed, we can
think of $F^2$ as an operator in an effective theory for soft gluons
where all short-distance degrees of freedom at scale
$Q^2$ have been ``integrated out", and there are no on-shell
partons to produce collinear singularities.

We can now imagine separating the full integral for $C_0$ into
the sum of two terms, one in which no gluons have vanishing momentum 
and another in which we group the neighborhoods of all
the points $k=0$.  For a generic diagram this would be an
arduous task, but one that can clearly be accomplished
in a finite number of steps.
The schematic result is
\bea
C_0=C_0^{({\rm reg})}+ {C_0^{({\rm pinch})}} + {{\cal O}(Q^{-6})}\, ,
\eea
where in the second, ``pinch" term, Mueller showed that to
the relative power $Q^{-6}$ all the
logarithms can be incorporated into the running coupling evaluated
at the gluon ``mass",
\bea
C_0^{({\rm pinch})}
 &=&  H(Q) {\int_0^{\kappa^2} dk^2 k^2\;  \alpha_s(k^2)} \nonumber \\
&=& H(Q) {\int_0^{\kappa^2} dk^2 k^2
\ {\alpha_s(Q^2) \over 1 + \left ({\alpha_s(Q^2)\over 4\pi}\right ) \beta_0\ln(k^2/Q^2)}}
 \nonumber\\
 &=& H(Q) {\int_0^{\kappa^2} dk^2 k^2\;
{4\pi\over \beta_0 \ln(k^2/\Lambda^2)}}\, ,
\label{irintegral}
\eea
with $\kappa\ll Q$ a new factorization scale, which we
introduce to separate these ``soft corners" of momentum space.
The function $H(Q)$ represents the short-distance-dominated factors
left over (the shaded region in Fig.\ \ref{opepinch}(a)).
We will call  such a reorganization of perturbation
theory  an ``internal resummation" \cite{gsth2002}, one
in which all logarithmic behavior in $\int f(\as(k^2))$ has
been absorbed into the running coupling.  
We also see that, because the perturbative running
coupling is undefined at the QCD scale $\Lambda$,
an internal resummation alone does not produce
a well-defined, finite result.

 Now, here's the operator product expansion:
soft integrals in $C_0^{({\rm pinch})}$ 
and for the perturbative expansion
$\langle 0| F^2 |0 \rangle $ are identical, something
that we can see by comparison of (a) and (b) in
Fig.\ \ref{opepinch} once we realize that the factor
of $k^2$ near the pinch surface corresponds to
the local vertex $F^2$.   
At this point we invoke what I like to call an
``axiom of substitution", also known as ``matching".
That is, we postulate that the true behavior of the two-point
function is found by systematically removing 
the soft corners of integrals in perturbation theory,
and replacing them with power corrections in the form
of matrix elements that have the same integrals in the same soft corners.
As we have argued, such a process will be consistent
with the operator product expansion (in this case),
and we write
\bea
{C_{\rm PT}\over Q^4} \rightarrow C_{\rm PT}^{\rm reg}\left (\frac{Q^2}{\mu^2},
\frac{\kappa}{Q},\alpha_s(Q)\right )
+C_{F^2}(Q,\kappa)\ {{\alpha_s \langle0|F^2(0)|0\rangle(\kappa) \over Q^4}}\, .
\eea

In summary, the nonconvergence of 
perturbation theory implies the need for
new (infrared) regularization.
The cost of introducing such a 
regularization is that the theory
requires a new nonperturbative parameter ({\it i.e}., $\langle F^2 \rangle$).
By the same token, the benefit
of this procedure is to recognize the
necessity of this same parameter ({\it i.e}., $\langle F^2 \rangle$),
which emerges from perturbation theory.
Such an analysis is possible, and indeed necessary,
whenever there is an internal resummation,
an integral of the form
$\int f(\alpha_s(k^2)) dk^2$ to all logs, for some funnction $f$
of the running coupling.

Of course, we may start from an alternative
viewpoint, assuming the OPE from the beginning.
In this case, we interpret the matching above as
a cancellation of high-order corrections between
leading-power perturbation theory and the 
operator matrix elements.
Whichever way we interpret this correspondence,
the actual  values of nonperturbative parameters 
will depend on the definition of perturbation theory.
The method of effective charges, discussed by
Maxwell \cite{effcharge} at this meeting,
shows how flexible the values of these 
parameters can be!

\begin{figure}[h]
\hbox{\hskip 2.0 in \epsfxsize=9cm \epsffile{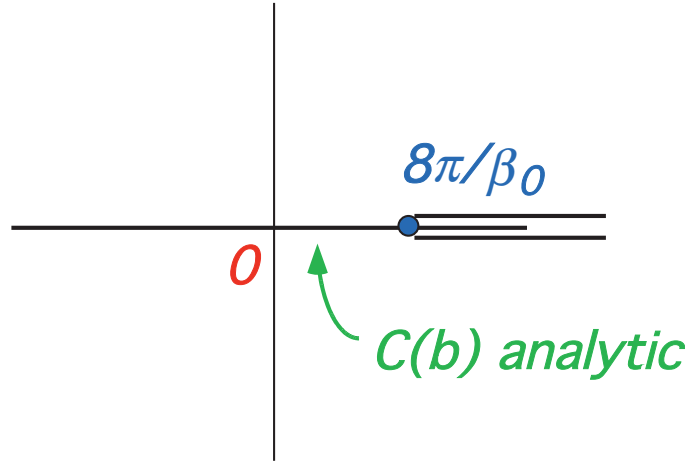}}
\caption{The Borel plane interpretation of Eqs.\ (\ref{irintegral})
and (\ref{borelint}). \label{borelplane}}
\end{figure}

Yet another common method is in terms of a Borel transform \cite{mueller85,thooft74},
which we can  see emerge from our analysis above.  
To do so, we may change variables in Eq.\ (\ref{irintegral}),
\bea
b\equiv 2\alpha_s(Q^2)\ln(Q^2/k^2)\, ,
\eea
in terms of which $C_0$ becomes
\bea
 C_0^{\rm pinch}\left (\alpha_s(Q)\right ) = 
{1\over 2}\; H(Q) Q^4\; \int_0^\infty db\; e^{-b/\alpha_s(Q)}
{1\over 1-{\beta_0\over 8\pi}\; b}\, ,
\label{borelint}
\eea
illustrated in Fig.\ \ref{borelplane}.
The explicit pole in the integrand is none other than the pole
in the perturbative running coupling.  Here, it should be
regarded as an ambiguity in the integral at the value
$b=8\pi/\beta_0$ of the transform variable.
Any finite redefinition of the integral to make it unambiguous
comes in only at the level of $Q^{-4}$ relative to the leading
contribution from the transform.

It is certainly worth noting that such ``infrared renormalon"
power corrections are not the only possible sources of
nonperturbative corrections.  The very existence of
solutions to the classical equations of motion,
such as instantons, implies nonconvergent behavior
of the perturbative series \cite{solutions}.  Another potential source
of nonperturbative corrections
are  ultraviolet renormalons,  poles on the negative
real axis of the $b$-plane \cite{uvrenorm}.   
While these corrections may well exist, 
 phenomenological evidence appears 
 at present to be associated  with the
 infrared effects, and in the remainder of this talk I will restrict my discussion
 to the latter.  

\section{Semi-inclusive cross sections}

The ideas described above took a big step forward in the mid-1990s,
with applications to semi-inclusive cross sections.  
The operator product expansion applies only to a very
small set of observables, but there is a much larger
class of observables that are infrared safe, that is,
calculable order-by-order in perturbation theory
with finite coefficients.  As for the example above,
we don't expect the resulting series to be
convergent, but once again we may ask whether
that nonconvergence could be set to our advantage.
My references below to the work that grew out
of this realization will be of necessity incomplete,
but at this point I can mention two very useful reviews, by
Beneke and Braun  \cite{BBpowerreview} and 
by Dasgupta and Salam \cite{DSeventreview}.

\subsection{Internal resummation and power corrections}

For many infrared safe quantities, we can
find a resummation of 
logarithms, which can be generated from
an integral over the running coupling, appearing
in a form similar to (\ref{irintegral}) above,
$(1/Q^2)\int_0^{Q^2} f(\alpha_s(k^2)) dk^2$,  for some $f$
and  some pertubative scale $Q$. 
Expanded back into a power series
in a fixed coupling like $\as(Q)$, all
integrals are finite, which is just a restatement
that such an expression is consistent with
infrared safety. 
In the light of our discussion above, however, we 
may be tempted to interpret the ambiguity in defining
the lower end of such an
integral, $0\ge k^2\le \Lambda_{\rm QCD}$, as signaling the 
presence of a nonperturbative power correction.
The approach just described has indeed been pursed by many
investigators for many 
infrared safe processes \cite{earlypower,Webber94,NasonSeymour95}.  In some sense it
is the simplest viewpoint to take, and one may
apply it wherever the running coupling is encountered
in an integral.   As we shall see shortly, it is a
particularly natural approach for resummed cross sections.

At the same time, to increase confidence in our 
interpretation, we might want to require that
such an expression organize ``all logs",
that is, in our terminology above, that it be
an  ``internal resummation".  
I will come back to this later.

\subsection{Massive gluons and dispersive couplings}

A very influential line of reasoning 
takes a viewpoint related to internal resummation, with a focus on the ``massive
gluon" introduced above.
At least one current of this reasoning  began with the observation that
 particle masses
give generic $m/Q$ corrections to event shapes (see below).
 For the event shape thrust this observation  goes back at least to R.\ Basu in 1984,
 for the specific case of the b quark \cite{Basu84}. 
In 1994, Webber \cite{Webber94} proposed using the gluon ``mass" 
in event shapes, generalizing
the work of Bigi {\it et al}. in heavy quark physics \cite{Bigietal}.
 This approach in turn
 can be given precise realization by ``large-$\beta_0$" or ``dressed gluon" \cite{dga,Berger04} 
 approximations,
which employ specific models for gluon self-energy diagrams.  
These and related ideas
remain quite useful in organizing power correcctions. 

In a sense, the apotheosis of the massive gluon
is found in the conjecture of Dokshitzer,
Marchesini and Webber (DMW) that the QCD coupling has a ``dispersive"
structure, conveniently written 
in terms of functions $\rho$ and $\alpha_{\rm eff}$ as \cite{dispersive}
\bea
\as(k^2) = - \int_0^\infty \frac{d\mu^2}{\mu^2+k^2}\ \rho_s(\mu^2) \, ,
\quad 
\rho_s(\mu^2) = \frac{d}{d\ln \mu^2}\alpha_{\rm eff}(\mu^2)\, .
\label{asdisper}
\eea
Applied to QED, the  function $\rho(\mu^2)$ is just the gauge-invariant photon self-energy,
which is essentially the QED perturbative running coupling.
The generalization to QCD in (\ref{asdisper}) defines
a formalism that can be implemented at fixed order, but whose extension
to all orders has not, to my knowledge, been fully explored.   
This apparent limitation, however, may not be that
restrictive, especially if one takes the viewpoint that $\as(\mu^2)$
itself behaves much more smoothly at low scales than  the exptrapolation of
its perturbative description \cite{smoothas}.   

With this possibility in mind, DMW proposed 
a template for any infrared safe observable $F$, which can be implemented with an NLO analysis.  
Consider, then, an arbitrary process dressed by a single gluon
that carries loop momentum  $k$, as in the example above.  After an 
integration by parts, $F$ can be put in the form
\bea
F(x,Q) = - \int_0^\infty \frac{d\mu^2}{\mu^2}\, \alpha_{\rm eff}(\mu)\, \frac{d{\cal F}(x,Q,\mu)}{d\mu^2}\, ,
\label{disperF}
\eea
with the function ${\cal F}(x,Q,\mu)$ the NLO underlying short-distance sub-process, and 
$\mu$ the variable ``mass" of the gluon that emerges from this 
subprocess.  For this expression to be meaningful,
it is only necessary that ${\cal F}$ remain finite for $\mu^2\to 0$,
and an expansion in $\mu/Q$ organizes power
corrections, much as for the example from $\rm e^+e^-$ annihilation
above.   

The next step is to approximate the ``effective" running coupling
$\alpha_{\rm eff}$ in (\ref{disperF}) as the sum
perturbative and nonperturbative terms.  Naturally, the  definition of
the latter depends on the former.   The very simplest perturbative running
coupling is a fixed coupling, say $\as(Q^2)$, dressed by one or two
terms from the beta function.  
In event shapes, discussed below, certain NLO effects
are naturally taken into account in a ``MC" or ``physical" scheme
for the coupling, related to the familiar $\overline{\rm MS}$ scheme
by
\bea
\as = \alpha_s^{\rm \overline{MS}}\left (1 + \frac{\as}{2\pi}\, K\right)\, , \quad K= C_A\left(\frac{67}{18} - \frac{\pi^2}{6}\right)  - \frac{5}{9}n_f\, .
\label{mcas}
\eea
In this way, the Landau pole is simply
discarded.   The idea is that well above  the
Landau pole (say  at the one GeV level), the perturbative running coupling should be 
replaced by a smooth nonperturbative function, that DMW call
$\delta\alpha_{\rm eff}$.  This function is {\it a priori}
unknown.  In Eq.\ (\ref{disperF}), however,
it will appear in  universal moments, such as
\bea
A_{2p} = \frac{C_F}{2\pi} \int_0^\infty \frac{d\mu^2}{\mu^2}\ \mu^{2p}\, 
\delta\alpha_{\rm eff}(\mu^2)\, ,
\eea
where the powers (and sometimes logarithms) of the ``gluon mass", $\mu$, come from
the expansion of ${\cal F}$.  Phenomenological studies of power corrections
then make it possible to determine  these moments and, by comparing different
processes, to  check their universality.
This procedure has been applied to many
infrared safe observables.  For inclusive processes, it
reproduces the operator product expansion  (with perhaps
some extra assumptions about the vanishing of certain
of the moments $A_q$), but its most important implications
are for semi-inclusive processes, to which the operator
product expansion does not directly apply.

\subsection{Event Shapes and the Milan Factor}

Let's recall that event shapes, $e(N)$ 
are numbers that depend on energy flow in final states
\cite{eflow}, $N$, and
define weighted cross sections, as
\bea
\frac{d\sigma}{de}
=
\sum_N \sigma(N) \; \delta\left( e(N)-e\right)\, ,
\eea
where $\sigma(N)$ is the cross section for state $N$.
When expanded to any fixed order, these cross sections
are generally infrared divergent and are {\it not} positive definite.
Because the event shapes are constructed to depend on
energy flow only, they assign the same weight to 
states that differ by the emission of zero-momentum partons,
and/or collinear rearrangements of partons.  If the function
$e(N)$ is sufficiently smooth, this is enough to ensure infrared safety
\cite{gs79}.

As measured in the long and distinguished history
of $\rm e^+e^-$ colliders, event shapes provide many tests for
the ideas described above, and much of this analysis has
concentrated on the DMW formalism.  There is a technical problem,
however, that was realized early on \cite{NasonSeymour95}.
By definition, event shapes weight different parts of phase space differently,
while the derivation of a dispersive coupling as in (\ref{asdisper}) from the
imaginary part of the gluon self-energy requires an 
inclusive sum with equal weight for all final states.
The ``MC" scheme mentioned above then doesn't get the entire $\as^2$ part right.
Applications of the dispersive approach 
to event shapes thus require a bit more analysis.

For a wide class of event shapes, however, it was realized that
within the assumption that low orders in $\alpha_{\rm eff}$
are adequate, it is possible to
compensate for the lack of complete inclusivity 
by simply rescaling the  $1/Q$ power correction by what has come to be
called the ``Milan factor" \cite{milanfac},
\bea
\frac{1}{Q} \rightarrow {\cal M}\, \frac{1}{Q}\, .
\label{Milanfac}
\eea
The Milan factor takes into account the mismatch between the inclusive
sum over soft gluon radiation in the dispersive coupling and 
the weighted sum in  event  shapes.
The factor is computable at fixed order, and 
is universal for a surprisingly wide variety of
observables.   Once again, it reflects a philosophy of  
fixed order for the nonperturbative as well as perturbative
coupling.  Thus,  the effect of two-loop corrections is taken
to be competitive to one-loop, but yet higher loops 
are assumed to be safely neglected.  

This set of ideas and applications was adopted
by many of the LEP experiments for their analyses,
and, after considerable care,  the basic concepts have been shown to
be quite useful and phenomenologically successful.
Of particular interest are applications to 
average event shapes, as summarized for example in Ref.\ \cite{DSeventreview}
and in the talk by Salam at this meeting.  Also, armed with this
formalism, the analysis has been extended to
event shapes with more than two jets and
with hadrons in the initial state \cite{extenddisper}.

\section{Resummation and Shape Functions}

Electron-positron data are available not only for average
values of event shapes, but also for their full distributions \cite{DSeventreview}.
We will discuss here primarily event shapes that
are constructed to vanish when the final state 
consists of two perfectly-collimated jets.  Qualitatively,
these distributions are well-described by fixed-order perturbation
theory only far from the two-jet limit.   At the same time, the 
bulk of the cross section is found quite near this limit.
(This is amply illustrated in Fig.\ \ref{hjmfig} below.)
As a result, the distributions of event shapes in general require perturbative
resummations.   In addition their phenomenological description normally
requires important  additional nonperturbative
input.   Many of the underlying results go back to the work
of Collins and Soper on two-particle correlations in $\rm e^+e^-$ annihilation,
with important implications to electroweak annihilation \cite{cs81,qtresum}.  

In the following, I will concentrate on a set of two-jet event shapes that were
recently introduced, the ``angularities".   Partly, this is because of
personal familiarity, but partly it is because the angularities were
introduced to make available a set of observables that depended
on an adjustable parameter.   This makes them amenable both
to theoretical and experimental analysis.

\subsection{Resummmed angularities in $\rm e^+e^-$ annihilation}

 The angularities \cite{Berger04}
 are a set of event shapes defined as
 \cite{Berger04,bks03,Banfi05,Berger03}
 \bea
\tau_a= \sum_{i\, {\rm in}\, N} 
\, {E_i\over Q}\, \left(\sin\theta_i\right)^a\, \left(1-|\cos\theta_i|\right)^{1-a}
=  \sum_{i\, {\rm in}\, N} \frac{p_{T,i}}{Q}\, e^{-(1-a)|\eta_i |}\, ,
\eea
where the sum is over particles $i$ of energy $E_i$ that appear in the final state,
and where  $\theta_i$ and $\eta_i$ are respectively the particle  angles 
 and rapidities defined relative to thrust axis.  
 
 The classic shapes thrust, $T$ and jet broadening, $B$ are
 given by
 \bea
 T = 1 -\tau_0\, , \quad B = \tau_1\, .
 \eea
 The cross section at zero angularity goes over to the total
 cross section in the limit $a\rightarrow\infty$, where
 all states have the same vanishing weight.
 The angularities thus interpolate between a 
 fairly extensive set of observables, all of which are
 available in the same data set.

Near the two-jet, $\tau_a=0$, limit, the angularities all
develop large singular corrections of the generic plus-distribution  type,
$\as^n [\ln^{2n-1}\tau_a/\tau_a]_+$ at $n$th order.   To derive
a self-consistent cross section for $\tau_a\to 0$, it is necessary
to resum these logarithms.  In the standard form 
familiar from thrust \cite{thrustresum}, the resummed cross
section for the entire set of angularities is 
given to next-to-leading logarithm (NLL) by 
an inverse Laplace transform with transform variable
$\nu$ \cite{Berger03}, proportional to the Born Cross section, $\sigma_{\rm tot}^{(0)}$,
\bea
\ {\sigma}
\left(\tau_a,Q,a \right) \!
&=& \sigma_{\rm tot}^{(0)}\; \int_C d\nu\, {\rm e}^{\nu\, \tau_a}\; 
\left[\, { J_{i}(\nu,p_{Ji})}\, \right]^2\, .
\label{rapinverse}
\eea
As usual, the contour $C$ may be taken parallel
to the imaginary axis in the $\nu$ complex plane.
The functions $J_i$ are themselves Laplace transforms
of factorized functions that describe the two jets, and whose
logarithmic $\nu$-dependence exponentiates,
\bea
{ J_{i}({\nu,p_{Ji})}}
&=&
\int_0 d\tau_a\, {\rm e}^{-\nu\tau_{Ji}}\; 
{ J_{i}({\tau_{Ji},p_{Ji})}}= {\rm e}^{{1\over 2}E(\nu,Q,a)}\, .
\label{Jtrans}
\eea
The leading logarithmic (LL) behavior of the exponent is $\as\ln^2\nu$,
which produces $\as^n\ln^{2n}\nu$ at $n$th order in
the transformed jets and hence in the transform of the cross section. 
Applying the inverse 
Laplace transform, we derive the leading, 
$\as^n [\ln^{2n-1}\tau_a/\tau_a]_+$, terms at each order of the cross section.

Once we go to NLL, we find that the  exponent is
again an integral over scales in the running coupling
 for all of the angularities,
\bea
E(\nu,Q,a) &=&
  2\, \int\limits_0^1 \frac{d u}{u} \Bigg[ \,
      \int\limits_{u^2 Q^2}^{u Q^2} \frac{d p_T^2}{p_T^2}
A\left(\as(p_T)\right)
      \left( {\rm e}^{- u^{1-a} \nu \left(p_T/Q\right)^{a} }-1 \right)
%\nonumber \\
   %   & &      
      + \frac{1}{2} B\left(\as(\sqrt{u} Q)\right) \left( {\rm e}^{-u
\left(\nu/2\right)^{2/(2-a)} } -1 \right)  \Bigg] \, .
      \nonumber\\
\label{Eform}
\eea
Again, the argument of $\alpha_s$ can vanish when $u$ and hence $p_T$ vanish,
yet an expansion in $\alpha_s(Q)$ is finite at any fixed order.
Thus the exponent itself has the nonconvergent form that
we have associated with power corrections above.  The
anomalous dimensions that enter at NLL are familiar,
and when expanded as
$A(\as) = \sum_{n=1}^\infty A^{(n)}\ \left({\as\over
\pi}\right)^n$,
and similarly for $B$, they are 
\bea
A^{(1)}   =  C_F
, \quad
A^{(2)}   = \frac{1}{2} C_F \, K \quad
B^{(1)} =  - \frac{3}{2} \, C_F\, ,
\eea
where $K$ is given as above in (\ref{mcas})
and where $A(\as)$ is the same anomalous dimension that appears
in the DGLAP evolution kernels. Thus at NLO, the anomalous dimension
$A(\as)$ is the same as the running coupling in the MC scheme.
 It should be noted, however
that beyond NLL, additional anomalous dimensions enter
these expressions, so that in (\ref{Eform}) we do not yet have
a complete internal resummation in the sense of organizing
all logarithmic corrections.

\subsection{Exponentiating power corrections, shape functions}

By studying the ambiguities of perturbation theory in the
spirit of the discussions above, one finds  that 
the first power correction in the moment space exponent
is of the form $\nu \lambda/Q$, with $\lambda$ a new
nonperturbative parameter.  The effect
of such a contribution is to shift the resummed
perturbative distribution \cite{shift}. 

For large $\nu$, the correction $\nu \lambda/Q$ may actually reach
unity, and we may expect that all powers of this dimensionless
quantity are competitive with the perturbative, leading-power exponent.
Shape functions \cite{eventshapes}, first introduced for the
thrust distribution, are an attempt to organize all such corrections into
a nonperturbative function.  Let's see how this works for the angularities.

Starting from the resummed exponent of (\ref{Eform}),
we introduce a new factorization scale $\kappa$, and 
decide to trust perturbation theory for $p_T>\kappa$.
We will call this contribution $E_{\rm  pert}(\nu,Q,\kappa,a)$.
For $p_T<\kappa$, on the other hand, we will
interpret integrals over the running coupling in terms
of new nonperturbative parameters as above.  

Following these steps, we exchange the order of the $u$
and $p_T$ integrals in (\ref{Eform}) and 
expand the exponentials in the $u$ integrands.  At each order of the
expansion, we can trivially perform the $u$ integral,
and we find
\bea
E(\nu,Q,a)     &=&   {E_{\rm  pert}(\nu,Q,\kappa,a)}
+\, \frac{2}{1-a}\
\sum_{n=1}^\infty\ \frac{1}{n\, n!}\, {\left(-{\nu\over Q}\right)}^{n}
\! \int\limits_{0}^{\kappa^2} {dp_T^2\over p_T^2}\; p_T^n\;
A\left(\alpha_s(p_T)\right)  + \dots
\nonumber\\
&\equiv& 
{E_{\rm  pert}(\nu,Q,\kappa,a)}
+ {\ln \tilde f_{a,{\rm NP}}\left(\frac{\nu}{Q},\kappa \right)} + \dots\, ,
\label{shapesum}
\eea
where in the first equality we have isolated all terms that
are powers of $\nu/Q$.  Neglected contributions, including
all those from the anomalous dimension $B(\as)$, are suppressed
by an additional power of $Q$ (when $a<1$).  In the second line, the sum
of all such terms is represented as the logarithm of an as-yet unknown 
``shape function" $f_a$.  Although undetermined, $f_a$ is a function
of the overall momentum scale $Q$ only through the combination
$\nu/Q$.

As is clear from Eq.\ (\ref{shapesum}),
the logarithm of the shape function enters additively
in the exponent, and therefore multiplicatively in
the moments of the cross
section. Inverting the moments, we therefore find a convolution,
\bea
\frac{d\sigma(\tau_a,Q)}{d\tau_a} =  \int d\xi\ f_{a,{\rm NP}}(\xi)\ 
\frac{d\sigma_{\rm PT}(\tau_a-\xi,Q)}{d\tau_a}\, .
\label{convol}
\eea
In momentum space, the shape function depends only on
the convolution variable, and  not on the momentum scale $Q$.
This suggests a strategy analogous to the fitting of parton
distribution functions, as first advocated in \cite{eventshapes}
for the thrust.
Looking at data from LEP,
one fits the shape function
at $Q=M_{\rm Z}$, where the data is
most plentiful.  Once the shape function
is in hand, it is only necessary to evaluate 
(\ref{convol}) to derive  predictions for all $Q$.

\begin{figure}[h]
\centerline{\epsfxsize=9cm \epsffile{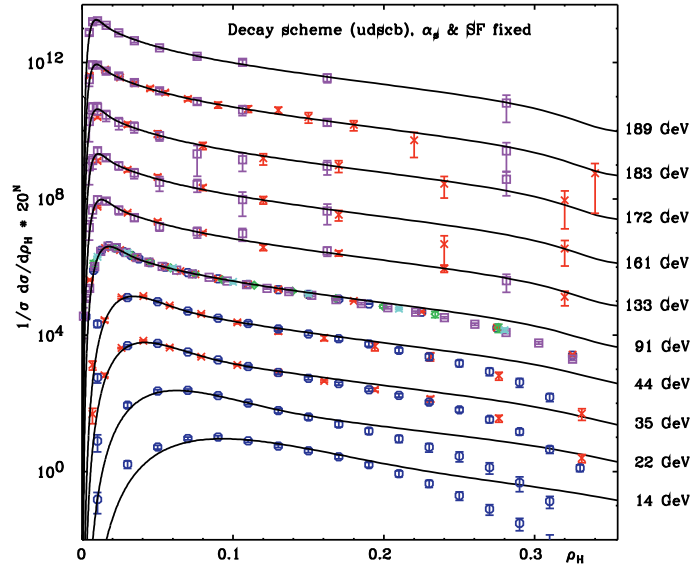}}
\caption{Event shape function fits for the heavy jet mass \cite{heavyjet},
from Gardi and Rathsman,  Ref.\ \cite{shapephenom}.
\label{hjmfig}}
\end{figure}

The phenomenology of shape functions for event shapes
has been extensively studied in Refs.\ \cite{shapephenom}
for the thrust and related observables.  
The impressive
quality of the results are illustrated  for
the ``heavy jet mass" \cite{heavyjet} in Fig.\ \ref{hjmfig}.

We have already observed that all such IR safe event shapes are related to correlations of 
nonlocal operators that describe energy flow \cite{eflow}.  
In Ref.\ \cite{Belitsky01}, the physical interpretation of
the shape functions was studied in this light.
Here it was argued that such considerations
suggest a simple functional form for the thrust  ($a=0$) shape function,
\bea
f_{0,\rm NP}(\rho) = {\rm const}\ \rho^{\alpha-1}\; {\rm e}^{-\beta\rho^2}\, ,
\eea
where the parameter
$\alpha$ is related to the number of particles produced per unit\ rapidity,
while $\beta$ measures correlations between the energy flow
in the two hemispheres centered about the jet directions.

\subsection{Scaling for the angularities}

Although the angularities for arbitrary parameter $a<1$ can  be resummed
to NLL in much the same way as the thrust and allied
$a=0$ event shapes, they have not yet been confronted with
experiment.  Such a comparison is arguably important,
however, because of an approximate
 scaling property that is implicit in the NLL
 discussion given above.  From Eq.\ (\ref{shapesum}),
 we observe that all terms in the expansion
 in $\nu/Q$ are proportional to $1/(1-a)$, which
 is the only $a$-dependence in the entire function
at the level of approximation where all (and only)
powers of $\nu/Q$ are taken into account,
\bea
\hspace{-30mm}
\ln \tilde f_{a,{\rm NP}}\left(\frac{\nu}{Q},\kappa \right)
=  \frac{1}{1-a}\; \sum_{n=1}^\infty\ \lambda_n(\kappa)
{\left(-{\nu\over Q}\right)}^{n}\, .
\label{faoneminusa}
\eea
This immediately gives the relation \cite{Berger03,Berger04}
\bea
\tilde f_{a}\left(\frac{\nu}{Q},\kappa\right)
=
\left[\, \tilde f_{0}\left(\frac{\nu}{Q},\kappa\right)\, 
\right]^{1\over 1-a}\, ,
\label{scaling}
\eea
which implies that we can start with the thrust shape
function at $a=0$ and predict shape functions for any
$a$.  Certainly it would be of great interest to see
if we can infer nonperturbative properties of the cross section in this manner.
The assumptions that go into this derivation of the scaling relation were 
studied in \cite{Berger03}, where it was argued
that the scaling tests the rapidity-independence of
hadronization from jets.  In addition, the powers shown
in Eq.\ (\ref{faoneminusa}) are formally leading
only for $a<1$, while the angularities are actually
infrared safe for all $a<2$.   The crossover between
competing power corrections has been treated in
Refs.\ \cite{Berger04,Banfi05}.  In any case,
a result either way would be of significance.  
 
\begin{figure}[h]
\centerline{\epsfxsize=9cm\epsfig{file=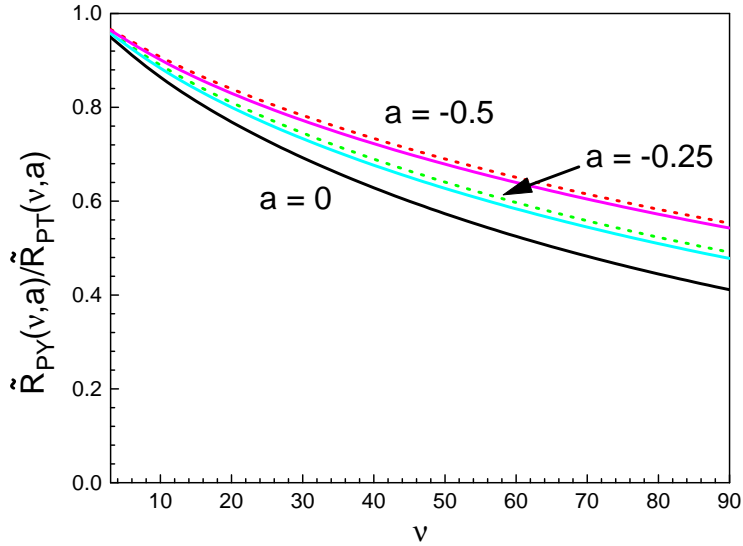,height=10cm,angle=270}}
\caption{Ratios of moments of angularity cross sections computed
from PYTHIA (${\rm \tilde R_{PY}}$) to the resummed perturbation theory moments
(${\rm \tilde R_{PT}}$) for $a=0,\ -0.5,\ -0.5$. the lines for the two latter
cases are the predictions of scaling. \label{scalingfig}}
\end{figure}

Some evidence for the relevance of the scaling
may be gleaned from PYTHIA, which can stand
in for the data analysis that we lack.  The result of such a comparison \cite{Berger03}
is shown in Fig.\ \ref{scalingfig}.   The relative success seen in
this plot certainly does not ensure a similar result
for actual data.   If the data did not show this behavior, however,
it would  say something about the hadronization
model built into the event generator.  And that is part of the motivation.   

Tests of the angularities are only one example of
the potential of $\rm e^+e^-$ event structure to address new questions
in the transition from perturbative to nonperturbative 
dynamics.  Most of the event shapes that were
used in the analyses of and since the 1990s were invented for the jet
physics of the late 1970s.  In the intervening time, we have learned 
how to ask a whole new set of questions, which existing
data may be able to answer for us.

\section{Beyond leading orders and leading powers}

While the shape function approach purports to organize
the entire set of nonperturbative powers in $\nu/Q$,
much of the original discussions \cite{eventshapes}, including
those on angularities \cite{Berger03,Berger04} were
based on NLL resummations.   As we have observed
above, it is also interesting to search for more complete treatments,
which can organize perturbation theory to arbitrary logarithmic
orders.    Such a treatment can be based on the factorization of
soft and collinear dynamics in gauge theories, as
already observed in \cite{eventshapes}.   Related results
have now been derived in the context of  soft-colllinear
effective theory \cite{Manohar03}, and I hope it will be fruitful
to study the relationship between these two methods.
[In fact, some steps in this direction were taken at this
workshop, and are described in \cite{Lee06}.]
In closing, I will briefly discuss the factorization-based formalism,
and try to give some sense of how it can enable us to
establish very general properties in perturbation theory.

\subsection{Operators and soft gluon emission}

\begin{figure}[h]
%\vbox{\vskip 1 true in}
\centerline{\epsfxsize=6cm \epsffile{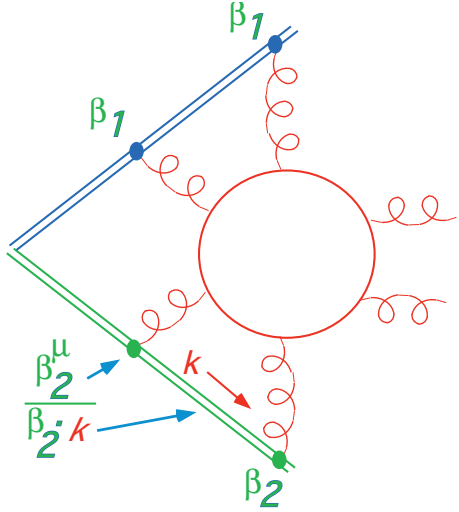}}
\caption{Typical (virtual) diagram in the eikonal cross section.\label{eikfig}}
\end{figure}

A basic result in factorization analysis is that the coupling of soft gluons
to jets is equivalent to their coupling to light-like ordered
exponentials, or Wilson lines,
\bea
\Phi^{(f)}_{\beta }(\lambda,x)
=
{\cal P}\exp \left (-ig\int_0^\lambda d\lambda' \beta \cdot A^{(f)}(\lambda'\beta+x)
\right )\, ,
\eea
where the field $A^{(f)}$ is in the matrix representation corresponding to 
the flavor $f$ of the parton that initiates the jet (quark or antiquark
for two-jet events in $\rm e^ +e^-$ annihilation).  Once this factorization
is carried out, we can define an event shape distribution from the
soft gluons to two-jet event shapes $e$,
\bea
{{d \sigma^{({\rm eik})}(e,\{\beta _i\})\over de}}
=
\sum_N\ {\delta\left(e-e(N)\right)}\ 
 \langle 0  |W^\dagger \, 
 {|N\rangle\, \langle N |}
\, W\,  | 0  \rangle\, ,
\label{eikS}
\eea
where $W$ is a product of Wilson lines for the quark and
antiquark,
\bea
W=  T\Bigg( \Phi^{(q)}_{\beta _q}(\infty,0)\, \Phi^{(\bar{q})}_{\beta_{\bar{q}}}(\infty,0)  
\Bigg )\, .
\eea
The perturbative expansion for these matrix elements
is equivalent to the eikonal approximation for radiation
from a light-like quark pair, illustrated in Fig.\ \ref{eikfig}.
The matrix elements above also correspond 
 to the leading operators in the soft-collinear effective theory of Ref.\ \cite{Manohar03}.

In  the complete cross section, the soft gluon distribution $d \sigma^{({\rm eik})}/de$,
Eq.\ (\ref{eikS}),
will appear in convolution with two functions describing the 
collinear dynamics of the jets.  The full cross section therefore factorizes into a simple
product in transform space.  Nonperturbative corrections of the jet functions 
should appear as powers in $1/m_J^2$, corresponding to powers of $\nu/Q^2$
in the transform space, which are therefore subleading compared to
those from the soft function.  A very important point is that the
factorization itself becomes better and better for smaller and smaller 
event shape values, that is, for narrower and narrower jets.
Corrections to factorization vanish as powers of $m_J/Q$,
and therefore decay as powers of $\nu$ in transform space.

\subsection{Eikonal exponentiation for event shapes}

The wonderful thing about eikonal cross sections like (\ref{eikS}) is
that they exponentiate algebraically, in terms of functions
whose perturbative properties are simpler than those of
the full theory.
For shape functions, the results are
equivalent to Eq.\ (\ref{Eform}) but more general.

\begin{figure}[b]
%\vbox{\vskip 1 true in}
\centerline{\epsfxsize=6cm \epsffile{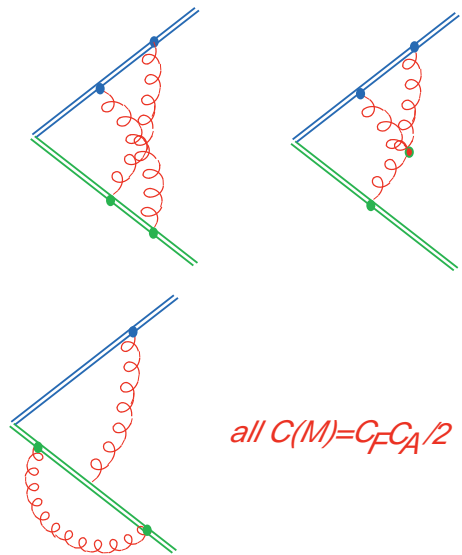}}
\caption{Examples of virtual web diagrams.  The webs
are irreducible under cuts of two eikonal lines.  All
of these webs have the same color factor, which in the full
theory is the color factor corresponding to the web with
the three-gluon vertex. \label{webfig}}
\end{figure}

The relevant functions have been called ``webs" \cite{webs}.
The web functions can be written as sums over 
 individual cut diagrams, ${\cal M}^2_N$.  A cut diagram is the union of
 a diagrammatic contribution to the amplitude for the production of state $N$,
 with another for the 
 complex conjugate amplitude.
The function ${\cal M}^2_N$ includes all diagrams of this type that cannot be
disconnected by cutting only two eikonal lines.
Each diagrammatic contribution to ${\cal M}^2_N$ is associated with a specific, overall
modified color factor, $C({\cal M}^2_N)$.  The web function at fixed
event shape $e$ can then be written formally as
\bea
E(e) = \sum_{{\rm states}\ N}\ \delta\left(e-e(N)\right)\,
 \sum_{\cal M} C({\cal M}^2_N)\; {\cal M}^2_N(e)\, .
\eea
Examples of virtual  web diagrams  are shown 
in Fig.\ \ref{webfig}.  In this set all have the same modified color factor.

As long as the event shape $e$ can be written as the sum over
the contributions of individual final-state particles, the eikonal cross
section at fixed total event shape $e$ is a convolution
of the form,
\bea
{{d \sigma^{({\rm eik})}(e)\over de}} = \sum_{n=0}^\infty\; {1\over n!}\; 
 \prod_{i=1}^n\; \int de_i\, E(e_i)\; \delta(e-\sum_i e_i) \, .
\eea
It therefore exponentiates under a Laplace transform,
\bea
\tilde S_e(\nu,Q) \equiv
\int_0 de\;  {\rm e}^{-\nu e}\ 
{d \sigma^{({\rm eik})}\over de} = \exp\, \left[ \int_0 de'\; {\rm e}^{-\nu e'}\; E(e')\, \right] \, ,
\eea
where dependence on the upper limit is exponentially suppressed for large $\nu$,
corresponding to $e\to 0$.

For light-like eikonal lines, the web functions themselves are
 boost and renormalization group invariant.
 To get a sense of how these properties can be used, consider
 a trivial event shape, in which the radiated energy
 is fixed.   By itself, such a cross section would have
 collinear singularitites, but these can be systematically 
 subtracted.  Indeed this is just what happens in
 a Drell-Yan cross section when the eikonal
 lines are incoming, or for a two-particle correlation \cite{cs81}
 when the eikonal lines are outgoing.  In either case,
 the eikonal annihilation cross section takes the form \cite{joint}
\bea
\hspace{-30mm}
\ln \tilde S_{\rm incl}(\nu,Q) &=&  \sum_{{\rm states}\ N}\ 
 \sum_{\cal M} C({\cal M}^2_N)\; {\cal M}^2_N(E_N/Q)\ e^{-\nu E_N/Q}
\nonumber\\
&=& \int_0^{Q^2} \frac{\rho(\as(u,\varepsilon))}{u^2}\;
\left[ K_0\left(\frac{2\nu u}{Q}\right) + \ln\frac{u}{Q}\right] 
+ 2\ln\bar \nu \int_0^{Q^2} \frac{du^2}{u^2} A(\as(u,\varepsilon))\, ,
\label{inclexp}
\eea
where boost and renormalization group invariance implies that the spectral function
$\rho(\as)$ can be written as a function of the running $\as$ only.  
Here $\bar{\nu}\equiv \nu e^{-\gamma_E}$ with $\gamma_E$
the Euler constant, and the integrals 
are defined by dimensional regularization in $D=4-2\varepsilon$
dimensions.

The final term in Eq.\ (\ref{inclexp})  represents the collinear subtractions, and
for such inclusive cross sections, the relevant factorization theorems actually
{\it require} a 
generalization of the ``MC-scheme" mentioned above, relating
the spectral function to the anomalous dimension $A(\as)$ to all
orders in perturbation theory, with a correction that begins at next-to-next to 
leading logarithm,
\bea
\rho(\as(u,\varepsilon)) = 2A(\as(u,\varepsilon)) + \frac{\partial D(\as,\varepsilon)}{\partial \ln \mu^2}\, .
\eea
In four dimensions, 
this correction,  $D(\as,0)$ is the all-orders generalization
of the NNLL ``D-term", long known in the resummed Drell-Yan
cross section \cite{Dterms}.   

Equation (\ref{inclexp}) is a true internally resummed expression.
The very specific form of the term in square brackets,
involving the Bessel function, follows from boost invariance
along the two-jet axis, and is accurate up to corrections that
decay exponenetially in $\nu$.  Expanded in powers of 
$\nu u/Q$, this predicts the form of power corrections (even powers only),
by the standard reasoning above.

We also learn from (\ref{inclexp}) that at NNLL the MC scheme, and
presumably its dispersive property, inherits what
appear to be process-dependent corrections.
The presence of such corrections, however, should allow us to test ideas like the freezing
of the coupling as we learn to apply these methods to
more general classes of observables, including event shapes.

\section{Conclusion: Looking Toward the Far Infrared}

The phenomenology of power corrections, especially in event shapes,
is one of the success stories of the past generation of high energy colliders.
Moving beyond the current stage of phenomenology, however, to
a better understanding of the physics will require both new ideas
and accessibility for data sets that can be used to test them.

Looking at just the few topics touched on above, it is clear that a fuller
phenomenology waits on systematic insights into formally-nonleading corrections
that appear at the level of powers of $\nu/Q^2$ in many event shapes.
Such corrections are closely connected to hadronization within jets,
and hence to parton-hadron duality \cite{duality}, already familiar
from deep-inelastic scattering in the narrow-jet, that is $x \to 1$, limit.
This general question may also be related to the role of the nonleading power corrections in
the $a<1$ angularities mentioned below Eq.\ (\ref{faoneminusa}).

A central issue still in the background 
is the role of quarks.  Most of the analysis of resummation
and power corrections is driven
by gluons, but eventually, it's primarily quarks that characterize the
hadrons that we observe.  Finally, as hinted at above,
the study of power corrections should eventually overlap with
event generator concepts like clusters and string breaking.
It is my hope that this workshop will be a stimulus for
addressing these problems, and the many others that I've surely missed.

\begin{acknowledgments}
I would like to thank the organizers of the FRIF Workshop for the opportunity
to speak at this exciting workshop, and to many of the participants, including 
but only, Volodya Braun,
Yuri Dokshitzer, Mrinal Dasgupta, Einan Gardi, Georges Grunberg, Klaus Hamacher,
Stefan Kluth, Gregory Korchemsky, Christopher Lee, 
Lorenzo Magnea, Chris Maxwell, Al Mueller, Gavin
Salam and Giulia Zanderighi
for so many illuminating conversations before, during and since the workshop.
Important points in Sec.\ V reflect ongoing work with Werner Vogelsang.
This work
was supported in part by the National Science Foundation, grants PHY-0354776 and PHY-0345922.
\end{acknowledgments}

\end{document}